\documentstyle[prl,aps,preprint]{revtex}

\begin{document}
\title{Lectures on Strings, D-branes and Gauge Theories\thanks{Lectures 
delivered by the first author in
the {\it Third Workshop
on Gravitation and Mathematical Physics,} Nov. 28-Dec. 3 1999, Le\'on Gto. M\'exico.}}
\author{Hugo Garc\'{\i}a-Compe\'an\thanks{
Email address: compean@fis.cinvestav.mx} and Oscar Loaiza-Brito\thanks{
Email address: oloaiza@fis.cinvestav.mx}}
\address{ Departamento de F\'{\i}sica\\
Centro de Investigaci\'on y de Estudios Avanzados del IPN\\
Apdo. Postal 14-740, 07000, M\'exico D.F., M\'exico}
\date{\today}
\maketitle

\begin{abstract}
\vskip-1.4truecm 

In these lectures we review the basic ideas of perturbative and
non-perturbative string theory. On the non-perturbative side we 
give an introduction to D-branes and string duality. 
The elementary concepts of non-BPS branes and noncommutative gauge theories
are also discussed.

\end{abstract}
\vskip 6truecm

\noindent
hep-th/0003019, CINVESTAV-FIS-00/13

\vskip -.5truecm





\newpage

\section{Introduction}

\setcounter{equation}{0}

String theory is by now, beyond the standard model of particle physics, the best and the
most sensible understanding of all the matter and their interactions
in an unified scheme. There are well known the
`esthetical' problems arising in the heart of the standard model of particles, such as:
the abundance of free parameters, the origin of flavor and the gauge group, etc. It is also generally
accepted that these problems require to be answered. Thus the standard model can be seen as
the low energy effective theory of a more fundamental theory which can solve the mentioned problems. 
It is also clear that the quantum
mechanics and general relativity cannot be reconciliated in the context of a perturbative
quantum
field theory of point particles. Hence the nonrenormalizability of the general relativity can
be seen as a genuine evidence that it is just an effective field theory and new physics
associated to the fast degrees of freedom should exist at higher energies. String theory
propose that these fast degrees of freedom are precisely the strings at the perturbative
level and at the non-perturbative level the relevant degrees of freedom are, in addition to the 
strings,  
higher-dimensional extended
objects called D-branes (dual degrees of freedom).
 The study of theories involving D-branes is just in the
starting stage and many surprises surely are coming up. Thus we are still at an exploratory
stage of the whole structure of the string theory. Therefore the theory is far to be completed and 
we cannot give yet concrete physical predictions to take 
contact with collider experiments and/or astrophysical observations. However many aspects of
theoretic character, necessary in order to make of string theory a physical theory, are quickly
in progress. The purpose of these lectures are to overview the basic ideas to understand these
progresses. This paper is 
an extended version of the lectures presented at the {\it Third Workshop
on Gravitation and Mathematical Physics} at Le\'on Gto. M\'exico. We don't pretend to be
exhaustive and we will limit ourselves to describe some basic elements of string theory
and some particular new developments as: non-BPS branes and noncommutative gauge theories. 
We apologize for omiting numerous original references and we prefer to cite review articles
and some few seminal papers.

In Sec. II we overview the string and the superstring theories from the perturbative
point of view. T-duality and  $D$-branes is 
considered in Sec. III. The Sec. IV is devoted to describe the string dualities and the web
of string theories connected by duality. M and F theories are also briefly described. Sec. V is devoted 
to review the non-BPS branes and their description in terms of topological K-Theory. Some recent
results by Witten and Moore-Witten, about the classification of Ramond-Ramond fields is also described.
The relation of string theory to noncommutative Yang-Mills theory and deformation
quantization theory is the theme of the Sec. VI.

\vskip 2truecm
\section{Perturbative String and Superstring Theories}

In this section we overview some basic aspects of bosonics and fermionic 
strings. We focus mainly in the description of the spectrum of the theory
in the light-cone gauge and the brief description of spectrum of 
the five consistent superstring theories (for details and further
developments see for instance \cite{gsw,polbook,lt,hat,oy,kir1}).

First of all consider, as usual, the action of a relativistic point
particle. It is given by  $S = -m \int
d{\tau} \sqrt{- \dot{X}^{\mu}\dot{X}_{\mu}}$, where $X^{\mu}$ are $D$ functions
representing the coordinates of the $(D-1,1)$-dimensional Minkowski spacetime
(the target space), $\dot{X}^{\mu} \equiv {d X^{\mu} \over d \tau}$ and  
$m$ can be identified with the mass of
the point
particle. This action is
proportional
to the length of the world-line of the relativistic particle.
In analogy with the relativistic point particle, the action describing
the dynamics of a string (one-dimensional object) moving in a 
$(D-1,1)$-dimensional Minkowski spacetime (the target space) is
proportional to the area ${\bf A}$ of the worldsheet. We know from the
theory of surfaces that such an area is given by ${\bf A}=\int \sqrt{det(-g)}$,
where $g$ is the induced metric (with signature $(-,+)$) on the worldsheet. The 
background  metric will be
denoted by $\eta_{\mu \nu}$ and  $\sigma^a=(\tau,\sigma)$  with $a=0,1$ are the local coordinates on the
worldsheet. $\eta_{\mu \nu}$ and $g_{ab}$ are related by  $g_{ab}= \eta_{\mu \nu}
{\partial}_a X^{\mu} {\partial}_b X^{\nu}$ with $\mu=0,1, \dots , D-1$. 
Thus the classical action
of a
relativistic string is given by the Nambu-Goto action

\begin{equation}
S_{NG}[X^{\mu}]=-{1\over 2 \pi {\alpha}'} \int d\tau d\sigma
\sqrt{-det({\partial}_a X^{\mu} {\partial}_b X^{\nu})},
\end{equation}
where $X^{\mu}$ are $D$ embedding functions of the worldsheet into
the target space $X$. Now introduce a metric $h$ describing the worldsheet geometry, 
we get a classically  equivalent action to the Nambu-Goto action. This is the Polyakov action

\begin{equation}
S_P[X^{\mu},h_{ab}] = - {1\over 4 \pi {\alpha}'} \int d^2 \sigma
\sqrt{-h}h^{ab}\partial_a X^{\mu} \partial_b X^{\nu} \eta_{\mu \nu},
\end{equation}
where the $X^{\mu}$'s are $D$ scalar fields on the worldsheet. Such a fields can be interpreted
as the coordinates of spacetime $X$ (target space), $h$ = det$({h}^{ab})$ and
$h_{ab}={\partial}_aX^{\mu}{\partial}_bX^{\nu} \eta_{\mu \nu}$.

Polyakov action has the following symmetries: $(i)$ Poincar\'e invariance,
$(ii)$ Worldsheet diffeomorphism invariance, and $(iii)$ Weyl invariance
(rescaling invariance).  The energy-momentum tensor of the two-dimensional
theory is given by

\begin{equation}
T^{ab}:= {1 \over \sqrt{- h}} {\delta S_P \over \delta h_{ab}}
={1\over 4 \pi {\alpha}'} \bigg({\partial}^a X^{\mu}{\partial}^b X_{\mu}-
{1\over 2}{h}^{ab} h^{cd} {\partial}_c X^{\mu}
{\partial}_d X_{\mu} \bigg) = 0.
\end{equation}

Invariance under worldsheet diffeomorphisms implies that it should be conserved {\it
i.e.}
${\nabla}_aT^{ab}=0$, while the Weyl invariance gives the traceless condition,
$T^a_a=0$. The equation of motion associated with Polyakov action is given by

\begin{equation}
\partial_a \bigg( \sqrt{-h} h^{ab} \partial_b X^{\mu} \bigg) = 0.
\end{equation}
Whose solutions should satisfy the boundary conditions for the open string: 
${\partial}_{\sigma}X^{\mu} {\mid}^{\ell=\pi}_0=0$ (Neumann) and for the closed
string: $X^{\mu} (\tau , \sigma )=X^{\mu}(\tau , \sigma + 2 \pi)$ (Dirichlet). Here $\ell=\pi$ is the 
characteristic
length of the open string. The variation of $S_P$ with respect to $h^{ab}$ leads
to the constraint equations $T_{ab} = 0$.
From now on we will work in the {\it conformal gauge}. In this gauge: $h_{ab} = \eta_{ab}$
and equations of motion (4) reduces to the Laplace equation whose solutions can be written 
as linear superposition of plane waves.

\newpage

\vskip 1truecm

\noindent
{\it 2.1 The Closed String} 

For the closed string the boundary condition $X^{\mu}(\tau,  \sigma )= X^{\mu}(\tau , \sigma +
2\pi )$, leads to the general solution of Eq. (4)

\begin{equation}
X^{\mu} = X^{\mu}_0 + {1 \over \pi T} P^{\mu}\tau + {i \over 2\sqrt{\pi T}} \sum_{n \neq 0}
{1\over n} \bigg\{ \alpha^{\mu}_n exp\bigg(-i2n(\tau - \sigma
)\bigg) + \tilde{\alpha}^{\mu}_n exp\bigg(-i2n(\tau + \sigma )\bigg)\bigg\}
\end{equation}
where $X^{\mu}_0$ and $P^{\mu}$ are the position and momentum of the
center-of-mass of the string and $\alpha^{\mu}_n$ and
$\tilde{\alpha}^{\mu}_n$ satisfy the conditions
${\alpha}_n^{ \mu *}={\alpha}^{\mu}_{-n}$ (left-movers) and $
\tilde{\alpha}_n^{\mu *}=\tilde{\alpha}^{\mu}_{-n}$ (right-movers).

\vskip 1truecm
\noindent
{\it 2.2 The Open String} 

For the open string the respective boundary condition is ${\partial}_{\sigma}X^{\mu} {\mid}^{\ell =
\pi}_0=0 $ 
(this is the only boundary condition which is Lorentz invariant) and the solution
is given by

\begin{equation}
X^{\mu}(\tau , \sigma )= X^{\mu}_0 + {1 \over \pi T}P^{\mu}\tau +{i\over \sqrt{\pi T}}
\sum_{n \neq 0} {1 \over n} {\alpha}^{\mu}_nexp \big(-in\tau\big) \cos (n\sigma )
\end{equation}
with the condition ${\alpha}^{\mu}_n = \tilde{\alpha}^{\mu}_{-n}.$

\vskip 1truecm

\noindent
{\it 2.3 Quantization}

The quantization of the closed bosonic string can be carried over, as usual, by using the Dirac
prescription to the center-of-mass and oscillator variables in the form

$$
[X^{\mu}_0,P^{\nu}]=i{\eta}^{\mu \nu},$$
$$
[{\alpha}^{\mu}_m,{\alpha}^{\nu}_n]=
[\tilde{\alpha}^{\mu}_m,\tilde{\alpha}^{\nu}_n]=m{\delta}_{m+n,0}{\eta}^{\mu
\nu},
$$
\begin{equation}
[{\alpha}^{\mu}_m,\tilde{\alpha}^{\nu}_n]=0.
\end{equation}
One can identify $({\alpha}^{\mu}_n,\tilde{\alpha}^{\mu}_n)$
with the creation operators and the corresponding operators
$({\alpha}^{\mu}_{-n},\tilde{\alpha}^{\mu}_{-n})$ with
the annihilation ones. In order to specify the physical states we first denote the center 
of mass state given by $|0;P^{\mu}\rangle$. The vacuum state is defined by $
{\alpha}^{\mu}_m |0,P^{\mu} \rangle=0$ and $P^{\mu}_m |0,P^{\mu} \rangle
=p^{\mu}\mid 0,P^{\mu} \rangle$ and similar for the right movings. For the zero modes these states have
negative norm
(ghosts). However one can choice a suitable gauge where ghosts decouple from the Hilbert space
when $D=26$. This is the subject of the following subsection. 

\newpage

\vskip 1truecm
\noindent
{\it 2.4 Light-cone Quantization}

Now we turn out to work in the so called {\it light-cone gauge}. In this
gauge it is possible to solve explicitly the Virasoro constraints (3). This is done 
by removing the light-cone coordinates $X^{\pm} = {1\over \sqrt{2}}(X^0\pm
X^D)$ leaving only the transverse coordinates $X^i$ representing the physical 
degrees of freedom (with $i=1, 2, \dots , D-2$). In this gauge the Virasoro constraints (3) are
explicitly 
solved. Thus the independent variables are $(X_0^-,P^+,X^j_0,P^j,
\alpha_n^j, \tilde{\alpha}_n^j)$. Operators $\alpha_n^-$  and $\tilde{\alpha}_n^-$ 
can be written in terms of $\alpha^j_n$ and $\tilde{\alpha}^j_n$ respectively as 
follows: ${\alpha}^-_n={1\over \sqrt{2 \alpha
'}P^+}(\sum_{m= - \infty}^{\infty} :{\alpha}^i_{n-m}{\alpha}^i_m:-2A{\delta}_n)
$ and $\tilde{\alpha}_n^- = {1\over \sqrt{2 \alpha
'}P^+}(\sum_{m= - \infty}^{\infty} :\tilde{\alpha}^i_{n-m}
\tilde{\alpha}^i_m:-2A{\delta}_n$). For the
open string we get
${\alpha}^-_n={1\over 2\sqrt{2 \alpha '}P^+}(\sum_{m = -\infty}^{\infty} :
{\alpha}^i_{n-m} {\alpha}^i_m:-2A{\delta}_n)
$. Here $: \cdot : $ stands for the normal ordering.

In this gauge the Hamiltonian is given by

\begin{equation}
H={1\over 2}(P^i)^2+N-A \ \ \  {\rm  (open \ string),}
\ \ \ \ \
H=(P^i)^2+ N_L + N_R-2A \ \ \ {\rm (closed  \ string)}
\end{equation}
where $N$ is the operator number, $N_L = \sum_{m = - \infty}^{\infty} : \alpha_{-m} \alpha_m:$, and
$N_R = \sum_{m=- \infty}^{\infty}: \tilde{\alpha}_{-m} \tilde{\alpha}_m:.$ The
{\it mass-shell condition} is given by $ \alpha
' M^2= (N-A)$ (open string) and $\alpha ' M^2=2(N_L + N_R-2A)$ (closed string). For the open string,
Lorentz invariance implies that the first excited state is massless and therefore $A=1$. In the light-cone 
gauge $A$ takes the form $A = - {D-2 \over 2} \sum_{n=1}^{\infty} n$.
From the fact $\sum_{n=1}^{\infty}n^{-s}=\zeta(s),$ where $\zeta$
is the Riemann's zeta function (which converges for $s>1$ and has a unique analytic continuation at 
$s=-1$, where it takes the value $-{1\over 12}$) then  $A=-{D-2\over
24}$ and therefore  $D=26$.

\vskip 1truecm

\noindent
{\it 2.5 Spectrum of the Bosonic String}

\noindent
{\it Closed Strings}

The spectrum of the closed string can be obtained from the combination of the left-moving
states and the the right-moving ones. The ground state ($N_L =N_R =0$) is given by 
$\alpha 'M^2=-4$. That means that the ground state includes a tachyon. The first
excited state ($N_L=1=N_R$) is massless and it is given by
${\alpha}^{i}_{-1} \tilde{\alpha}^{j}_{-1} |0,P \rangle$. This state can be naturally 
decomposed into irreducible representations of the little group $SO(24)$ as follows

$$
{\alpha}^i_{-1}\tilde{\alpha}^j_{-1}\mid 0,P \rangle=
{\alpha}^{[i}_{-1}\tilde{\alpha}^{j]}_{-1}\mid
0,P \rangle + \bigg({\alpha}^{(i}_{-1}\tilde{\alpha}^{j)}_{-1}-{1\over D-2}{\delta}^{ij}
{\alpha}^k_{-1}\tilde{\alpha}^k_{-1}\bigg)\mid 
0,P \rangle$$
\begin{equation}
+~{1\over D-2}{\delta}^{ij}{\alpha}^k_{-1}\tilde{\alpha}^k_{-1}\mid 0,P \rangle .
\end{equation}
The first term of the rhs is interpreted as a spin 2 massless particle
$g_{ij}$ ({\it graviton}).  The second term is a range 2 anti-symmetric tensor
$B_{ij}$. While the last term is an scalar field 
$\phi$ ({\it dilaton}). Higher excited massive states are combinations of representations
of the little group SO$(25).$

\vskip 1truecm

\noindent
{\it Open Strings}

For the open string, the ground state includes once again a tachyon since 
$\alpha 'M^2=-1$. The first exited state 
$N=1$ is given by a massless vector field in 26 dimensions. The second excitation level
is given by the massive states ${\alpha}^{i}_{-2}\mid 0,P \rangle$ and
${\alpha}^{i}_{-1}{\alpha}^{j}_{-1}\mid 0,P \rangle$ which are in irreducible representations
of the little group SO$(25)$. 

\vskip 1truecm

\noindent
{\it 2.6 Superstrings}

In bosonic string theory there are two big problems. The first one is the presence
of tachyons in the spectrum. The second one is that there are no spacetime fermions.
Here is where superstrings come to the rescue. A superstring is described, despite of 
the usual bosonic fields $X^{\mu}$, by 
fermionic fields $\psi^{\mu} _{L,R}$ on the worldsheet. Which satisfy anticommutation rules and where
the $L$ and $R$ denote the left and right worldsheet chirality respectively. The action for the
superstring is given by

\begin{equation}
L_{SS}=-{1\over 8\pi}\int d^2 \sigma
\sqrt{-h} \bigg(h^{ab}\partial_aX^\mu\partial_bX_\mu +
2i \bar{\psi}^{\mu}\gamma^{a}\partial_{a}\psi_{\mu} -i \bar{\chi}_a\gamma^b
\gamma^a\psi^{\mu}\big(\partial_bX_\mu-{i\over 4}\bar{\chi}_b\psi_\mu \big)\bigg),
\end{equation}
where  $\psi^{\mu}$ and $\chi_a$ are the superpartners of $X^{\mu}$ and the tetrad field $e^a$, 
respectively.
In the superconformal gauge and in light-cone coordinates  it can be reduced to

\begin{equation}
L_{SS}= {1 \over 2 \pi}
\int \bigg(\partial_L X^{\mu}  \partial_R
X_{\mu} + i \psi^{\mu}_R\partial_L \psi_{\mu R} +
i \psi^{\mu}_L \partial_R \psi_{\mu R} \bigg).
\end{equation}

In analogy to the bosonic case, the local dynamics 
of the worldsheet metric is manifestly independent of quantum
corrections if the critical spacetime dimension $D$ is 10. Thus the string
oscillates in the 8 transverse dimensions. The action (10) is 
invariant under: $(i)$ worldsheet supersymmetry, $(ii)$ Weyl transformations,
$(iii)$ super-Weyl transformations, $(iv)$ Poincar\'e transformations and
$(v)$ Worldsheet reparametrizations. The equation of motion for the $X's$ fields is 
the same that in the bosonic case (Laplace equation) and 
whose general solution is given by Eqs. (5) or (6). Equation of
motion for the fermionic field is the Dirac equation in two dimensions.
Constraints here are more involved and they are called the {\it super-Virasoro
constraints}. However in the light-cone gauge, everything simplifies and 
the transverse coordinates (eight coordinates) become the bosonic physical degrees
of freedom together with their corresponding supersymmetric partners. Analogously 
to the bosonic case, 
massless states of the spectrum come into representations of the little group SO(8)
of SO$(9,1)$, while that the massive states lie into representations of the little group 
SO$(9)$.

For the closed string there are two possibilities for the boundary conditions of fermions:
$(i)$ periodic
boundary conditions (Ramond ({\bf R}) sector) 
$\psi^{\mu}_{L,R}(\sigma) = + \psi^{\mu}_{L,R}(\sigma + 2 \pi)$
and $(ii)$
anti-periodic boundary conditions
(Neveu-Schwarz ({\bf NS}) sector) $\psi^{\mu}_{L,R}(\sigma) = - \psi^{\mu}_{L,R}(\sigma + 2 \pi)$. 
Solutions of Dirac equation satisfying these boundary conditions are
\begin{equation}
\psi^{\mu}_L(\sigma,\tau) = \sum_n \bar{\psi}^{\mu}_{-n}exp\bigg(-in(\tau +\sigma  
)\bigg), \ \ \ \ \  \psi^{\mu}_R(\sigma,\tau) =
\sum_n \psi^{\mu}_{n}exp \bigg(-in(\tau -\sigma
)\bigg).
\end{equation}
In the case
of the fermions in the {\bf R} sector 
$n$ is integer and it is semi-integer in the {\bf NS} sector.

The quantization of the
superstring come from the promotion of the fields $X^{\mu}$ and $\psi^{\mu}$ to operators 
whose oscillator variables are operators satisfying the relations 
$[\alpha^\mu _n, \alpha^\nu _m]_-~= n\delta_{m+n,0}\eta^{\mu\nu}$ and 
$[\psi^\mu _n,\psi^\nu _m]_+~=~\eta^{\mu\nu}\delta_{m+n,0},$ where $[,]_-$ and 
$[,]_+$ stand for commutator and anti-commutator respectively.

The zero modes of $\alpha$ are diagonal in the Fock space and its
eigenvalue can be identified with its momentum. For the {\bf NS} sector there is no
fermionic zero modes but they can exist for the {\bf R} sector and they  satisfy a Clifford
algebra $[\psi^\mu _0,\psi^\nu _0]_+~=~\eta^{\mu\nu}$. The Hamiltonian for the
closed superstring is given by $H_{L,R}=N_{L,R}+ {1\over 2}P^2_{L,R}-A_{L,R}$. For the
{\bf NS} sector $A={1 \over 2}$, while for the {\bf R} sector $A=0$. The mass is given by
$M^2 = M_L^2 + M_R^2$ with  
${1\over 2} M^2_{L,R}=N_{L,R}-A_{L,R}$.

There are five consistent  superstring theories: Type IIA, IIB, Type I, SO(32) and
$E_8 \times E_8$ heterotic strings. In what follows of this section we briefly describe the spectrum in
each one of them.

\vskip 1truecm
\noindent
{\it 2.7 Type II Superstring Theories}

In this case the theory consist of closed strings only. They are theories with ${\cal N} =2$
spacetime supersymmetry. For this reason, there are 8 scalar fields (representing
the 8 transverse coordinates to the string) and one Weyl-Majorana spinor. There are 
8 left-moving and 8 right-moving fermions. 

In the {\bf NS} sector there is still a tachyon in the ground state.  But in the
supersymmetric case this problem can be solved through the introduction of
the called {\bf GSO} projection. This projection eliminates the tachyon in the {\bf NS} sector
and it acts in the {\bf R} sector as a ten-dimensional spacetime chirality operator. That means
that the 
application of the {\bf GSO} projection operator defines the chirality of a 
massless spinor in the {\bf R} sector. Thus from the left and right
moving sectors, one can construct states in
four different sectors: $(i)$ {\bf NS-NS},
$(ii)$ {\bf NS-R}, $(iii)$ {\bf R-NS} and $(iv)$ {\bf R-R}. Taking account the two types of
chirality $L$ and $R$ one has two possibilities:

\noindent
$a)-$ The {\bf GSO} projections on the left and right fermions produce different
chirality in the ground state of the {\bf R} sector ({\it Type IIA}).

\noindent
$b)-${\bf GSO} projection are equal in left and right sectors and the ground states 
in the {\bf R} sector, have the same chirality ({\it Type IIB}). Thus the spectrum for the Type IIA and
IIB superstring theories is:

\vskip 1truecm

\noindent
{\it Type IIA}

The {\bf NS-NS} sector has a symmetric tensor field $g_{\mu \nu}$ (spacetime metric), an antisymmetric
tensor field
$B_{\mu \nu}$ and a scalar field $\phi$ (dilaton). In the {\bf R-R} sector there is a vector field
$A_{\mu}$ associated with a 1-form $C_1$ ($A_{\mu} \Leftrightarrow C_1$) and a rank 3 totally  
antisymmetric tensor $C_{\mu \nu \rho} \Leftrightarrow C_3$. In general the {\bf R-R} sector
consist of $p$-forms $G_p= dC_{p-1}$ (where $C_{p}$ are called RR fields) on the ten-dimensional
spacetime $X$ with $p$
even
{\it i.e.}
$G_0$, $G_2$, $G_4$, \dots . In the {\bf NS-R} and {\bf R-NS} sectors we have two gravitinos with opposite
chirality and the supersymmetric partners of the mentioned bosonic fields.

\vskip 1truecm
\noindent
{\it Type IIB}

In the {\bf NS-NS} sector Type IIB theory has exactly the same spectrum that of Type IIA theory.
On the {\bf R-R} sector it has a scalar field $\chi \Leftrightarrow C_0$, an antisymmetric tensor field
$B'_{\mu \nu} \Leftrightarrow C_2$ and a rank 4 totally antisymmetric tensor $D_{\mu \nu \rho \sigma}
\Leftrightarrow C_4$
whose field strength is self-dual {\it i.e.},
$ G_5= d C_4$ with $*G_5 = +G_5$. Similar than for the case of Type IIA theory one has, in general, 
RR fields given by $p$-forms $G_p = d C_{p-1}$ on the spacetime $X$ with $p$ odd
{\it i.e.}
$G_1$, $G_3$, $G_5$, \dots . The {\bf NS-R} and {\bf R-NS} sectors do contain  gravitinos with the
same chirality and the corresponding fermionic matter.

\vskip 1truecm
\noindent
{\it 2.8 Type I Superstrings}

In this case the $L$ and $R$ degrees of freedom are the same. Type I and Type IIB theories have the
same spectrum, except that in the former one the states which are not invariant
under the change of orientation of the worldsheet, are projected out. This worldsheet parity
$\Omega$ interchanges
the left and right modes. Type I superstring theory is a theory of breakable closed strings, thus it
incorporates also open strings. The $\Omega$ operation leave invariant only one half of
the spacetime supersymmetry, thus the theory is ${\cal N}=1$.

The spectrum of bosonic massless states in the {\bf NS-NS} sector is:
$g_{\mu \nu}$ (spacetime metric) and $\phi$ (dilaton) from the closed sector and $B_{\mu \nu}$ is
projected
out.  On the {\bf R-R} sector there is an
antisymmetric field $B_{\mu \nu}$ of the closed sector. The open string sector is necessary in
order to cancel tadpole diagrams. A contribution to the spectrum come from this sector. Chan-Paton 
factors can be added at the boundaries of open strings. Hence the cancellation of the tadpole are 
needed 32 labels at each end. Therefore in the {\bf NS-NS} sector there are 496 gauge fields in the
adjoint representation of SO(32).

\vskip 1truecm

\noindent
{\it 2.9 Heterotic Superstrings}

This kind of theory involves only closed strings. Thus there are left and right sectors. The
left-moving sector contains a bosonic string theory and the right-moving sector contains
superstrings. This theory is supersymmetric on the right sector only, thus the theory contains
${\cal N}=1$ spacetime supersymmetry. The momentum at the left sector $P_L$ lives in 26 dimensions, while
$P_R$
lives in 10 dimensions. It is natural to identify the first ten components
of $P_L$ with $P_R$. Consistency of the theory tell us that the extra 16 dimensions should
belong to the root lattice $E_8 \times E_8$ or a ${\bf Z}_2$-sublattice of the SO(32) weight
lattice. 

The spectrum consists of a tachyon in the ground state of the left-moving sector. In both sectors
we have the spacetime metric $g_{\mu \nu}$, the antisymmetric tensor $B_{\mu \nu}$, the dilaton
$\phi$ and finally there are 496 gauge fields $A_{\mu}$ in the adjoint representation
of the gauge group $E_8 \times E_8$ or SO$(32)$.

\vskip 2truecm
\section{Toroidal Compactification, $T$-duality and D-branes}

D-branes are, despite of the dual fundamental degrees of freedom in string theory,
extremely interesting and useful tools to study nonperturbative  properties of 
string and field theories (for a classic review see \cite{pol}). Non-perturbative 
properties of supersymmetric gauge theories can be better understanding as the 
world-volume effective theory of some configurations of intersecting D-branes
(for a review see \cite{gk}). D-branes also are very important to connect gauge
theories with gravity. This is the starting point of the AdS/CFT correspondence 
or Maldacena's conjecture. We don't review this interesting subject in this paper,
however the reader can consult the excellent review \cite{malda}.
Roughly speaking D-branes are static solutions of string equations which satisfy 
Dirichlet boundary conditions. That means that open strings can end on them. 
To explain these objects we follow the traditional way, by using T-duality on
open strings we will see that Neumann conditions are turned out into the Dirichlet ones.
To motivate the subject we first consider T-duality in closed bosonic string theory.

\vskip 1truecm
\noindent
{\it 3.1 T-duality in Closed Strings}

The general solution of Eq. (4) in the conformal gauge can be written 
as $X^\mu (\sigma , \tau )=X^\mu_R(\sigma^-)+X^\mu_L (\sigma^+)$, where
$\sigma^{\pm}=\sigma \pm \tau$. Now, take one coordinate, say $X^{25}$ and compactify it on a 
circle of radius $R$. Thus
we have that $X^{25}$ can be identified with $X^{25} + 2\pi R m$ where $m$ is called the {\it winding
number}. The general solution for $X^{25}$ with the above compactification condition is

$$
X^{25}_R(\sigma^-)=X^{25}_{0R} +\sqrt{\frac{\alpha '}{2}}P^{25}_R(\tau -\sigma
)+i\sqrt{\frac{\alpha '}{2}}\sum_{l \neq 0}\frac{1}{l}\alpha^{25}_{R,l}exp
\bigg(-il(\tau - \sigma )\bigg)$$

\begin{equation}
X^{25}_L(\sigma^+)=X^{25}_{0L}+\sqrt{\frac{\alpha '}{2}}P^{25}_L(\tau +
\sigma)+i\sqrt{\frac{\alpha '}{2}}\sum_{n \neq
0}\frac{1}{l}\alpha^{25}_{L,l}exp\bigg(-il(\tau +\sigma )\bigg),
\end{equation}
where
\begin{equation}
P^{25}_{R, L}=\frac{1}{\sqrt{2}} \bigg(\frac{\sqrt{\alpha '}}{R}n \mp 
\frac{R}{\sqrt{\alpha '}}m\bigg).
\end{equation}
Here $n$ and $m$ are integers representing the discrete momentum and the winding
number, respectively. The latter has not analogous in field theory.
While the canonical momentum is given by $P^{25}=\frac{1}{\sqrt{2\alpha
'}}(P^{25}_L+P^{25}_R)$. Now, by the mass shell condition, the mass of the perturbative states 
is given by
$M^2=M_L^2+M_R^2$, with

\begin{equation}
M^2_{L,R}=-\frac{1}{2}P^\mu P_\mu =\frac{1}{2}(P^{25}_{L,R})^2+\frac{2}{\alpha '}(N_{L,R}-1).
\end{equation}

We can see that for all states with $m \neq 0$, as $R \rightarrow \infty$ the mass
become infinity, while  $m = 0$ implies that the states take all values for $n$ and form a
continuum. At the case when $R \rightarrow 0$, for states with $n \neq 0$, mass
become infinity. However in the limit $R \rightarrow 0$ for $n =0$ states with all
$m$ values produce a continuum in the spectrum. So, in this limit the compactified
dimension disappears. For this reason, we can say that the mass spectrum of the theories at radius $R$
and $\frac{\alpha '}{R}$ are identical when we interchange $n \Leftrightarrow m$. This
symmetry is known as  {\it T-duality}.

The importance of T-duality lies in the fact that the T-duality transformation is a parity
transformation acting on the left and right moving
degrees of freedom. It leaves invariant the left movers and changes the sign of the right movers
(see Eq. (14))
\begin{equation}
P^{25}_L \rightarrow P^{25}_L, \ \ \ \ \ \ \ \
P^{25}_R \rightarrow -P^{25}_R.
\end{equation}

The action of T-duality transformation must leave invariant the whole theory
(at all order in perturbation theory). Thus, all kind of interacting states in certain theory
should correspond to those states belonging to the dual theory. In this context,
also the vertex operators are invariant. For instance the tachyonic vertex operators are

\begin{equation}
V_L= exp(iP_L^{25}X^{25}_L), \ \ \ \ \ \ \ \ \
V_R= exp(iP^{25}_RX^{25}_R).
\end{equation}
Under T-duality, $X^{25}_L \rightarrow X^{25}_L$ and $X^{25}_R \rightarrow
-X^{25}_R$; and from the general solution Eq. (13), ${\alpha}^{25}_{R,i} \rightarrow
-{\alpha}^{25}_{R,i}$, $X^{25}_{0R} \rightarrow -X^{25}_{0R}$. Thus, T-duality
interchanges $n \Leftrightarrow m$ (Kaluza-Klein modes $\Leftrightarrow$ winding
number) and $R \Leftrightarrow {\alpha ' \over R}$ in closed string theory.

\vskip 1truecm

\noindent
{\it 3.2 T-duality in Open Strings}

Now, consider {\it open strings} with Neumann boundary conditions. Take again the
$25^{th}$ coordinate and compactify it on a circle of radius $R$, but keeping Neumann
conditions. As in the case of closed string, center of mass momentum takes only discrete
values $P^{25}=\frac{n}{R}$. While there is not analogous for the winding
number. So, when $R \rightarrow 0$ all states with nonzero momentum go to infinity
mass, and do not form a continuum. This behavior is similar as in field theory, but
now there is something new. The general solutions are 

$$
X^{25}_R =\frac{X_0^{25}}{2} -\frac{a}{2} + \alpha 'P^{25}(\tau -\sigma
) +i\sqrt{\frac{\alpha '}{2}}\sum_{l \neq 0}\frac{1}{l} \alpha^{25}_l exp \bigg(-i2l(\tau
-\sigma ) \bigg),$$

\begin{equation}
X^{25}_L =\frac{X_0^{25}}{2} +\frac{a}{2} +\alpha 'P^{25}(\tau +\sigma
)+i\sqrt{\frac{\alpha '}{2}}\sum_{l \neq 0}\frac{1}{l}\alpha^{25}_l exp \bigg(-i2l(\tau
+\sigma ) \bigg)
\end{equation}
where $a$ is a constant.
Thus,
$X^{25}(\sigma ,\tau ) = X^{25}_R(\sigma^-) +X^{25}_L(\sigma^+)=X_0^{25}
+\frac{2\alpha 'n}{R}\tau +  oscillator \ terms.
$
Taking the limit $R \rightarrow 0$, only the $n=0$ mode survives. Because of this,
the string seems to move in 25 spacetime dimensions. In other words, the strings vibrate in 24
transversal directions. T-duality provides a new T-dual coordinate defined by
$\tilde{X}^{25}(\sigma ,\tau
)=X^{25}_L(\sigma ,\tau )-X^{25}_R(\sigma ,\tau )$. Now, taking
$\tilde{R}=\frac{\alpha
'}{R}$ we have $\tilde{X}^{25}(\sigma ,\tau )=a +2 \tilde{R} \sigma n + oscillator \ terms.$
Using the boundary conditions at $\sigma =0,\pi$ one has
$ \tilde{X}^{25}(\sigma ,\tau ) \mid_{\sigma =0} = a$ and $
\tilde{X}^{25}(\sigma ,\tau )\mid_{\sigma =\pi}=a +2\pi \tilde{R}n.$			
Thus, we started with an open bosonic string theory with Neumann boundary
conditions, and T-duality and a compactification on a circle in the $25^{th}$
dimension, give us Dirichlet boundary conditions in such a coordinate. We can
visualize this saying that an open string has its endpoints fixed at a hyperplane
with 24 dimensions.

Strings with $n=0$ lie on a 24 dimensional plane space (D24-brane). Strings with
$n=1$ has one endpoint at a hyperplane and the other at a different hyperplane which
is separated from the first one by a factor equal to $2\pi \tilde{R}$, and so on. But
if we compactify $p$ of the $X^i$ directions over a $T^{p}$ torus ($i=1,...,p$). Thus, after T-dualizing
them we have strings with endpoints fixed at hyperplane with $25-p$ dimensions, the D$(25-p)$-brane. 

Summarizing: the system of open strings moving  freely in spacetime with $p$ compactified dimensions 
on $T^p$ is
equivalent, under T-duality, to strings whose enpoints are fixed at a D(25-p)-brane {\it i.e.} 
obeying Neumann boundary conditions in the $X^i$ longitudinal directions ($i=1,\dots ,p$) and
Dirichlet ones in the transverse coordinates $X^m$ ($m=p+1,...,25$).

The effect of T-dualizing a coordinate is to change the nature of the boundary conditions, from
Neumann  to Dirichlet and viceverse. If one dualize a longitudinal coordinate this coordinate will 
satisfies the Dirichlet condition and the D$p$-brane becomes a D$(p+1)$-brane. But if the dualized 
coordinate is one of the transverse coordinates the D$p$-brane becomes a D$(p-1)$-brane.

T-duality also acts conversely. We can think to begin with 
a closed string theory, and compactify it on to a circle in
the $25^{th}$ coordinate, and then by imposing Dirichlet conditions, obtain a
D-brane. This is precisely what occurs in Type II theory, a theory of closed strings.

\vskip 1truecm

\noindent
{\it Spectrum and Wilson Lines}

Now, we will see how emerges a gauge field on the D$p$-brane world-volume. Again, for the mass
shell condition for open bosonic strings and because T-duality 
$M^2=(\frac{n}{\alpha '}\tilde{R})^2+\frac{1}{\alpha '}(N-1)$. The massless state
($N=1$, $n=0$) implies that the gauge boson ${\alpha}^{\mu}_{-1}\mid 0 \rangle$ ($U(1)$
gauge boson) lies on to the D24-brane world-volume. On the other hand,
${\alpha}^{25}_{-1}\mid 0 \rangle$ has a {\it vev} (vacuum expectation value) which describes
the position $\tilde{X}^{25}$ of the D-brane after T-dualizing. Thus, we can say
in general, there is a gauge theory $U(1)$ over the world volume of the D$p$-brane.

Consider now an {\it orientable open string}. The endpoints of the string carry
charge under a non-Abelian gauge group. For Type II theories the gauge group is
$U(N)$. One endpoint transforms under the fundamental
representation {\bf N} of $U(N)$ and the other one, under its complex conjugate representation
(the anti-fundamental one) {\bf\=N}.

The ground state wave function is specified by the center of mass momentum and by
the charges of the endpoints. Thus implies the existence of a basis $\mid k; ij \rangle$
called {\it Chan-Paton basis}. States $\mid k;ij \rangle$ of the Chan-Paton basis are those
states which carry charge 1 under the $i^{th}$ $U(1)$ generator and $-1$ under the
$j^{th}$ $U(1)$ generator. So, we can decompose the wave function for ground state
as $\mid k;a \rangle=\sum_{i,j=1}^N \mid k;ij \rangle \lambda^a_{ij}$ where $\lambda^a_{ij}$ are
called {\it Chan-Paton factors}. From this, we see that it is possible to add degrees of freedom
to endpoints of the string, that are precisely the Chan-Paton factors.

This is consistent with the theory, because the Chan-Paton factors have a Hamiltonian
which do not posses dynamical structure. So, if one endpoint to the string is
prepared in a certain state, it always will remains the same. It can be deduced from
this, that $\lambda^a \longrightarrow U\lambda^aU^{-1}$ with $U\in$ $U(N)$. Thus, the worldsheet 
theory is
symmetric under $U(N)$, and this global symmetry is a gauge symmetry in spacetime. So the
vector state at massless level ${\alpha}^{\mu}_{-1}\mid k,a \rangle$ is a $U(N)$ gauge boson.

When we have a gauge configuration with non trivial line integral around a
compactified dimension (i.e a circle), we said there is a Wilson line. In case of
open strings with gauge group $U(N)$, a toroidal compactification of the $25^{th}$
dimension on a circle of radius $R$. If we choice a background field $A^{25}$ given
by $A^{25} =\frac{1}{2\pi R} diag(\theta_1,...,\theta_N)$ a Wilson line
appears. Moreover, if $\theta_i=0$, $i=1,...,l$ and $\theta_j \neq 0$, $j=l+1,...,N$
then gauge group is broken: $U(N) \longrightarrow U(l) \times U(1)^{N-l}$. It is
possible to deduce that $\theta_i$ plays the role of a Higgs field.
Because string states with Chan-Paton quantum numbers $\mid ij \rangle$ have charges $1$
under $i^{th}$ $U(1)$ factor (and $-1$ under $j^{th}$ $U(1)$ factor) and neutral with
all others; canonical momentum is given now by $
P^{25}_{(ij)} \Longrightarrow \frac{n}{R} + \frac{(\theta_j -\theta_i )}{2\pi R}.
$
Returning to the mass shell condition it results,

\begin{equation}
M^2_{ij}=\bigg(\frac{n}{R} +{\theta_j - \theta_i \over 2\pi R}\bigg)^2 +\frac{1}{\alpha '}(N-1).
\end{equation}

Massless states ($N=1, n=0$) are those in where $i=j$ (diagonal terms) or for which
$\theta_j=\theta_i$ $(i \neq j)$. Now, T-dualizing we have
$
\tilde{X}^{25}_{ij}(\sigma ,\tau )=a +(2n
+\frac{\theta_j- \theta_i}{\pi})\tilde{R}\sigma + oscillator \ terms.
$ Taking $a=\theta_i \tilde{R}$,
$
\tilde{X}^{25}_{ij}(0,\tau )= \theta_i \tilde{R}
$ and
$
\tilde{X}^{25}_{ij}(\pi ,\tau )= 2\pi n \tilde{R} + \theta_j \tilde{R}.
$
This give us a set of $N$ D-branes whose positions are given by
$\theta_j\tilde{R}$, and each set is separated from its initial positions
($\theta_j=0$) by a factor equal to $2\pi \tilde{R}$.
Open strings with both endpoints on the same D-brane gives massless gauge bosons.
The set of $N$ D-branes give us $U(1)^N$ gauge group. An open string with one endpoint 
in one D-brane, and the other endpoint in a different D-brane, yields a
massive state with $M \sim (\theta_j -\theta_i)\tilde{R}$. Mass decreases when two
different D-branes approximate to each other, and are null when become the same.
When all D-branes take up the same position, the gauge group is enhanced from $U(1)^N$ to 
$U(N).$ On the D-brane world-volume there are also scalar fields in the adjoint representation
of the gauge group $U(N)$. The scalars parametrize the transverse positions of the D-brane in the 
target space $X$. 

\vskip 1truecm

\noindent
{\it 3.3 D-brane Actions and Ramond-Ramond Charges}

\noindent
{\it D-Brane Action}

With the massless spectrum on the D-brane world-volume it is possible to construct a low
energy effective action. For open strings massless fields are interacting with the closed 
strings massless spectrum
from the {\bf NS-NS} sector. Let $\xi^a$ with $a=0, \dots , p$ the wold-volume
coordinates. The effective action is the gauge invariant action well known as the Born-Infeld action

\begin{equation}
S_D = - T_p \int_W d^{p+1} \xi e^{-\Phi}	\sqrt{ det\big( G_{ab} + B_{ab} + 2 \pi
\alpha ' F_{ab} \big)},
\end{equation}
where $T_p$ is the tension of the D-brane, $G_{ab}$ is the world-volume induced metric, $B_{ab}$ is the
induced antisymmetric
field, $F_{ab}$ is the Abelian field strength on $W$ and $\Phi$ is the dilaton field.

For $N$ D-branes the massless fields turns out to be $N \times N$ matrices and the action turns out to be
non-Abelian Bon-Infeld action (for a nice review about the Born-Infeld action in string theory see 
\cite{tseytlin})

\begin{equation}
S_D = - T_p \int_W d^{p+1} \xi e^{- \Phi} Tr\bigg(\sqrt{ det\big( G_{ab} + B_{ab}
+ 2 \pi \alpha ' F_{ab}\big)} + O\big( [X^m,X^n]^2 \big) \bigg)
\end{equation}
where $m,n = p+1, \dots , 10$. The scalar fields $X^m$ representing the transverse positions become
$N \times N$ matrices and so, the spacetime become a noncommutative spacetime. We will come back 
later to this interesting point.

\vskip 1truecm

\noindent
{\it Ramond-Ramond Charges}

D-branes are coupled to Ramond-Ramond (RR) fields $G_p$. The complete effective action on the D-brane
world-volume
$W$ which take into account this coupling is  

\begin{equation}
S_D = - T_p \int_W d^{p+1} \xi \bigg\{e^{-\Phi}	\sqrt{ det\big( G_{ab} + B_{ab} + 2 \pi
\alpha ' F_{ab} \big)} + i \mu_p \int_W \sum_p C_{(p+1)} Tr\bigg( e^{2 \pi \alpha ' (F+B)}\bigg) \bigg\}
\end{equation}
where $\mu_p$ us the RR charge. RR charges can be computed by considering the anomalous behavior
of the action at intersections of D-branes \cite{mm}. Thus RR charge is given by

\begin{equation}
Q_{RR} = ch(j!E) \sqrt{ \hat{A}(TX)}
\end{equation}
where $j : W \hookrightarrow X$. Here $E$ is the Chan-Paton bundle over $X$, 
$\hat{A}(TX)$ is the genus
of the spacetime manifold $X$. This
gives an ample evidence that the RR charges take values not in a cohomology theory, but in fact,
in a K-Theory \cite{mm}. This result was further developed by Witten in the context of non-BPS brane
configurations worked out by A. Sen. This subject will be reviewed below in Sec. V.

\vskip 2truecm
\section{String Duality}

\noindent
{\it 4.1 Duality in Field Theory}

Duality is a notion which in the last years has led to remarkable advances
in nonperturbative quantum field theory. It is an old known type of symmetry
which by interchanging the electric and magnetic fields leaves invariant
the vacuum Maxwell equations.  It was extended by Dirac to include
sources, with the well known price of the prediction of monopoles, which
appear as the dual particles to the electrically charged ones and whose
existence could not be confirmed up to now. Dirac obtained that the
couplings (charges) of the electrical and magnetical charged particles are
the inverse of each other, $i.e.$ as the electrical force is `weak', and
it can be treated perturbatively, then the magnetic force among monopoles
will be `strong' (for some reviews see \cite{peskin,kir2,quevedo}).

This duality, called `$S$-duality', has inspired a great deal of research
in the last years. By this means, many non-perturbative exact results have
been established. In particular, the exact Wilson effective action of
${\cal N}=2$ supersymmetric gauge theories has been computed by Seiberg and 
Witten, showing the
duality symmetries of these effective theories.  It turns out that the
dual description is quite adequate to address standard non-perturbative
problems of Yang-Mills theory, such as confinement, chiral symmetry
breaking, etc.

${\cal N}=4$ supersymmetric gauge theories in four dimensions have   
vanishing renormalization group $\beta$-function. Montonen and Olive
conjectured that (at the quantum level) these theories would possess an
SL$(2,{\bf Z})$ exact dual symmetry. Many evidences of this fact
have been found, although a rigorous proof does not exist at present.  For
${\cal N}=2$ supersymmetric gauge theories in four dimensions, the
$\beta$-function in general does not vanish. So, Montonen-Olive conjecture
cannot be longer valid in the same sense as for ${\cal N}=4$ theories.
However, Seiberg and Witten found
that a strong-weak coupling 'effective duality' can be defined on its low
energy effective theory for the cases pure and with matter. The
quantum moduli space of the pure theory is identified with a complex
plane, the $u$-plane, with singularities located at the points $ u=\pm  
1,\infty$. It turns out that at $u = \pm 1$ the original Yang-Mills theory
is strongly coupled, but effective duality permits the weak coupling
description at these points in terms of monopoles or dyons (dual variables).
${\cal N} =1$ gauge theories are also in the class of theories with
non-vanishing $\beta$-function. More general, for a gauge group SU$(N_c)$,
an effective non-Abelian duality is implemented even when the gauge
symmetry is unbroken. It has a non-Abelian Coulomb phase.  Seiberg has
shown that this non-Abelian Coulomb branch is dual to another non-Abelian
Coulomb branch of a theory with gauge group SU$( N_f - N_c)$, where $N_f$
is the number of flavors. ${\cal N}=1$ theories have a rich phase
structure. Thus, it seems that in supersymmetric gauge
theories strong-weak coupling duality can only be defined for some
particular phases.

For non-supersymmetric gauge theories in four dimensions, the subject of
duality has been explored recently in the Abelian as well as in the   
non-Abelian cases. In the
Abelian case (on a curved compact four-manifold $X$)  the $CP$ violating 
Maxwell theory partition function $Z(\tau)$, transforms as a {\it modular
form} under a finite index subgroup $\Gamma_0(2)$ of SL$(2,{\bf Z})$.  
The dependence parameter of the partition function 
is given by $ \tau = {\theta \over 2 \pi} + {4 \pi i \over e^2}$, where
$e$ is the Abelian coupling constant and $\theta$ is the usual theta
angle.  In the case of non-Abelian non-supersymmetric gauge theories,
strong-coupling dual theories can be constructed which results in a kind  
of dual ``massive" non-linear sigma models.  The starting Yang-Mills theory
contains a $CP$-violating $\theta$-term and it turns out to be equivalent
to the linear combination of the actions corresponding to the self-dual  
and anti-self-dual field strengths.

\vskip 1truecm
\noindent
{\it 4.2 String Duality}

In Sec. I we have described the massless spectrum  of the five consistent superstring theories in ten
dimensions. Additional theories can be 
constructed in lower dimensions by compactification of some of the ten dimensions.
Thus
the ten-dimensional spacetime $X$ looks like the product $X= K^d \times {\bf R}^{1,9-d}$, with
$K$ a suitable compact manifold or orbifold.
Depending on which compact space is taken, it will be the quantity of preserved supersymmetry. 

All five theories
and their compactifications are parametrized by: the string coupling constant $g_S$, the geometry 
of the compact manifold $K$, the topology of $K$ and the spectrum of bosonic fields in the {\bf NS-NS} 
and the {\bf R-R} sectors.
Thus one can define the {\it string moduli space} of each one  of the theories as the
space of all associated parameters. Moreover, it can
be defined a map between two of these moduli spaces. The dual map is defined as the map
${\cal S}: {\cal M} \to {\cal M}'$ between the moduli spaces ${\cal M}$ and ${\cal M}'$ such that 
the strong coupling region of ${\cal M}$ is interchanged with the weak-coupling region of 
${\cal M}'$ and viceverse. One can define another map ${\cal T}: {\cal M} \to {\cal M}'$ which
interchanges the volume $V$ of $K$ for ${1 \over V}$. One example of the map ${\cal T}$ is
the equivalence, by T-duality, between the theories Type IIA compactified on ${\bf S}^1$ at radius $R$ and
the Type IIB theory on ${\bf S}^1$ at raduis ${1 \over R}$.
The theories Het($E_8 \times E_8$) and Het(SO$(32)$) is another example. In this section we will
follows the Sen's review \cite{senlec}. Another useful reviews are \cite{schwarz,town,vafa,kir3}.

\vskip 1truecm

\noindent
{\it 4.3 Type I-SO(32)-Heterotic Duality}

In order to analyze the duality between Type I and SO(32) heterotic string theories we 
recall from Sec. II the spectrum of both theories. These fields are the dynamical fields of 
a supergravity Lagrangian in ten dimensions. 
Type I string theory has in the {\bf NS-NS} sector the fields: the metric $g_{\mu \nu}^I$, the dilaton
$\Phi^I$ and in the {\bf R-R} sector: the antisymmetric tensor $B_{\mu \nu}^I$. Also there are 496
gauge bosons $A^{aI}_{\mu}$ in the adjoint representation of the gauge group SO(32). For the SO(32)
heterotic 
string theory
the spectrum consist of: the spacetime metric $g_{\mu \nu}^H$, the dilaton field $\Phi^H$, the
antisymmetric
tensor $B_{\mu \nu}^H$ and 496 gauge fields $A^{aH}_{\mu}$ in the adjoint representation of SO(32). Both
theories 
have spacetime supersymmetry ${\cal N} = 1$. The effective action for the massless fields of the Type
I supergravity effective action $S_I$ is defined at tree-level on the disk. Thus the string coupling
constant $g_s^I$ arises in
the Einstein frame
as $exp (-\Phi^I /4)$. While the heterotic action $S_H$ is defined on the sphere and $g_S^H$ is given by
$exp (\Phi^H / 4)$. The comparison of these two actions in the Einstein frame leads to the
following identification of the fields

$$
 g_{\mu \nu}^I = g_{\mu \nu}^H, \ \ \ \ \ \ \ \  B_{\mu \nu}^I = B_{\mu \nu}^H 
$$

\begin{equation}
 A_{\mu}^{aI} = A_{\mu}^{aH}, \ \ \ \ \ \ \ \ \  \Phi^I = - \Phi^H.
\end{equation}
This give us many information, the first relation tell us that the metrics of both theories are the same.
The second relation interchanges the $B$ fields in the {\bf NS-NS} and the {\bf R-R} sectors. That 
interchanges
heterotic strings  by Type I D1-branes. The third  relation identifies the gauge fields coming from the
Chan-Paton factors from the Type I side with the gauge fields coming from the 16 compactified
internal dimensions of the heterotic string. Finally, the opposite sign for the dilaton relation means that
the string coupling constant $g^I_S$ is inverted $g^{H}_S = 1/ g^I_S$ within this identification,
and interchanges the strong and
weak couplings of both theories leading to the explicit realization of the ${\cal S}$ map. 

\vskip 1truecm

\noindent
{\it 4.4 Type II-Heterotic Duality}

Lower dimensional theories constructed up on compactification can have different spacetime 
supersymmetry. Thus it can be very useful to 
find dual pairs by compactifying two string theories with different spacetime supersymmetry
on different spaces $K$ in such a way that they become to have the same spacetime supersymmetry.

Perhaps the most famous example is the $S$-dual pair between the Type II theory on $K3$ and the
heterotic theory on $T^4$. To describe more generally these kind of dualities we first give some 
preliminaries. Let ${\cal A}$ and ${\cal B}$ two different theories of the family of string
theories. ${\cal A}$ and ${\cal B}$ are compactified on $K_{\cal A}$ and $K_{\cal B}$ respectively.
Consider the dual pair

\begin{equation}
{\cal A}/K_{\cal A} \Longleftrightarrow {\cal B}/K_{\cal B}
\end{equation}
then we can construct the more general dual pair 

\begin{equation}
{\cal A}/Q_{\cal A} \Longleftrightarrow {\cal B}/Q_{\cal B},
\end{equation}
where $K_{\cal A} - Q_{\cal A} \to D$ and $K_{\cal B} - Q_{\cal B} \to D$ are fibrations and
$D$ is an auxiliary finite dimensional manifold.

These insights are very useful to construct dual pairs for theories with eight supercharges.
An example of this is the pair in six dimensions with ${\cal A} = IIA$, $K_{\cal A}= K3$ and ${\cal
B} = Het$, $K_{\cal B} = T^4$ {\it i.e.}

\begin{equation}
IIA/K3 \Longleftrightarrow Het/T^4.
\end{equation} 
From this a dual pair can be constructed in four dimensions with the auxiliary space $D={\bf CP}^1$
being the complex projective space, thus we have 

\begin{equation}
IIA/CY  \Longleftrightarrow Het/K3 \times T^2,
\end{equation}
where $T^4 - Q_{IIA} \to {\bf CP}^1$ and $K3 - Q_{Het} \to {\bf CP}^1$ are fibrations. As
can be
observed the four-dimensional theories have ${\cal N}=2$ supersymmetry and the duality uses 
K3-fibrations.

\vskip 1truecm
\noindent
{\it 4.5 M-Theory}

We have described how to construct dual pairs of string theories. By the uses of the ${\cal S}$ and the 
${\cal T}$ maps a
network of theories can be constructed in various dimensions all of them related by dualities.
However new theories can emerge from this picture, this is the case of M-theory. M-theory
(the name come from `mystery', `magic', `matrix', `membrane', etc.) was
originally defined as the strong coupling limit for Type IIA string theory \cite{em}. 
At the effective low
energy action level, Type IIA theory is described by the Type IIA supergravity theory and it is 
known that this theory can be obtained from the dimensional reduction of the eleven dimensional
supergravity theory (a theory known from the 70's years). Let $Y$ be the eleven dimensional manifold,
taking $Y= X^{10} \times {\bf S}^1_R$
the compactification radius $R$ is proportional to $g_{10,10} \equiv \Phi$. Thus the limit 
$\Phi \to \infty$ corresponds to the limit $R \to \infty$ and thus the strong coupling limit of the
Type IIA theory corresponds to the 11 dimensional supergravity. It is conjectured that there exist
an eleven dimensional fundamental theory whose low energy limit is the 11 dimensional supergravity
theory. At the present time the degrees of freedom  are still unknown, through at the macroscopic level
they should be membranes and fivebranes (also called M-two-branes and M-fivebranes). There is a 
proposal to describe dof of M-theory in terms of a gas of D0-branes. This is  the called `Matrix Theory'.
This proposal as been quite successful (for some reviews see \cite{banks,taylor} and references therein).

\vskip 1truecm

\noindent
{\it 4.6 Horava-Witten Theory} 

Just as the M-theory compactification on ${\bf S}^1_R$ leads to the Type IIA theory, 
Horava and Witten realized that orbifold compactifications leads to the
$E_8 \times E_8$ heterotic theory in ten dimensions (see for instance \cite{town}). More precisely

\begin{equation}  
{\rm M}/{\bf S}^1/{\bf Z}_2 \Longleftrightarrow E_8 \times E_8 \ Het 
\end{equation}
where ${\bf S}^1/ {\bf Z}_2$ is homeomorphic to the finite interval $I$ and the $M$-theory is thus
defined on $Y = X^{10} \times I$. From the ten-dimensional point of view, this configuration is 
seen as two parallel planes placed at the two boundaries $\partial I$ of $I$. Dimensional reduction and
anomalies
cancellation conditions imply that the gauge degrees of freedom should be  
trapped on the ten-dimensional planes $X$ with the gauge group being $E_8$ in each plane.
While that the gravity is propagating in the bulk and thus both copies of $X$'s are only connected
gravitationally. 
 
\vskip 1truecm

\noindent
{\it 4.7 F-Theory}

$F$-Theory was formulated by C. Vafa, looking for an analog theory to M-Theory for
describing non-perturbative compactifications of Type IIB theory
(for a review see \cite{vafa,senlec}). Usually in perturbative
compactifications the parameter $\lambda = a + i exp(-\Phi/2)$ is taken to be constant. $F$-theory 
generalizes this fact by considering variable $\lambda$. Thus $F$-theory is defined as a 
twelve-dimensional theory whose compactification on the elliptic fibration
$T^2 - {\cal M} \to D$, gives the Type IIB theory compactified on $D$ (for a suitable
space $D$) with the identification of $\lambda(\vec{z})$ with the modulus $\tau(\vec{z})$ of the torus
$T^2$. These compactifications can be related to the $M$-theory compactifications
through the known $S$ mapping ${\cal S}: IIA \to M/{\bf S}^1$ and the ${\cal T}$ map
between Type IIA and IIB theories. This gives

\begin{equation}
F/{\cal M}\times {\bf S}^1 \Longleftrightarrow M/{\cal M}.
\end{equation}
Thus the spectrum of massless states of $F$-theory compactifications can be
described in terms of $M$-theory. Other interesting $F$-theory compactifications
are the Calabi-Yau compactifications

\begin{equation}
 F/CY \Leftrightarrow  Het/K3.
\end{equation}

\vskip 1truecm

\noindent
{\it 4.8 Gravitational Duality}

As a matter of fact, string theory constitutes nowadays the only
consistent and phenomenologically acceptable way to quantize gravity.  It
contains in its low energy limit Einstein gravity.  Thus, a legitimate  
question is the one of which is the `dual' theory of gravity or, more
precisely, how gravity behaves under duality transformations.

Gravitational analogs of non-perturbative gauge theories were studied
several years ago, particularly in the context of gravitational Bogomolny
bound. As recently was shown \cite{hull}, there are additional
non-standard $p$-branes in $D=10$ type II superstring theory and $D=11$
M-theory, and which are required by U-duality. These branes were termed
`gravitational branes' (`G-branes'), because they carry global charges   
which correspond to the $ADM$ momentum $P_M$ and to its `dual', a
$(D-5)$-form $K_{M_1 \cdots M_{D-5}},$ which is related to the NUT charge.
These charges are `dual' in the same sense that the electric and magnetic
charges are dual in Maxwell theory, but they appear in the purely
gravitational sector of the theory. Last year, Hull has shown in
\cite{hull} that these global charges $P$ and $K$ arise as central charges
of the supersymmetric algebra of type II superstring theory and
M-theory. Thus the complete spectrum of BPS states should include the
gravitational sector.

Finally, a different approach to the `gravitational duality' was worked out 
by using some techniques of strong-weak
coupling
duality for non-supersymmetric Yang-Mills theories were applied to the 
MacDowell-Mansouri dynamical gravity (for a review see \cite{cor}).  One would 
suspect that both approaches might be related 
in some sense. One could expect that the gauge theory of gravity would be
realized as the effective low energy theory on the `G-branes'.

\vskip 2truecm

\section{Non-BPS Branes and K-theory}

\noindent
{\it 5.1 Non-BPS Branes}

The notion of D-branes as BPS states implies the existence of certain supersymmetric theory
on the world-volume of the D-brane. However it is extremaly
relevant the consideration of non-supersymmetric theories (in order to describe our
non-supersymmetric world) and here is where it is important the construction of brane
configurations without remanent supersymmetry. A. Sen proposed the construction of such 
non-supersymmetric
configurations by considering pairs of D-branes and anti-D-branes (for a nice review see
\cite{senbps} see also \cite{lerda,nonbps}). These configurations 
break all supersymmetry and the spectrum on the world-volume has a tachyon which cannot be 
cancelled by {\bf GSO} projection. The presence of this tachyon leads to unstable brane configuration
and the configuration decay into an stable BPS configuration . The
classification of these stable D-branes was given by Witten in terms of topological K-Theory in the 
beautiful seminal paper \cite{ktheory} (for a review of this exciting subject see
\cite{olsen}).

In order to fix some notation let $X$ be the ten-dimensional spacetime manifold and let
$W$ be a $(p+1)$-dimensional submanifold of $X$.  Branes or antibranes or both together
can be wrapped on $W$. When configurations of $N$ coincident branes or antibranes only
are wrapped on $W$, the world-volume spectra on $W$ consists of a vector multiplet and
scalars in some representation of the gauge group. These configurations can be described
through Chan-Paton bundles which are U$(N)$ gauge bundles $E$ over $W$ for Type II
superstring theory and by SO$(N)$ or Sp$(N)$ bundles in Type I theory. Gauge fields from
the vector multiplet define a U$(N)$ gauge connection for Type II theory (or SO$(N)$ or
Sp$(N)$ gauge connection for Type I theory) on the (corresponding) Chan-Paton bundle.
{\bf GSO} projection cancels the usual tachyonic degrees of freedom. Something similar occurs
for the anti-brane sector.

The description of coincident $N_1$ coincident $p$-branes and $N_2$ $p$-anti-branes
wrapped on $W$ leads to the consideration pairs of gauge bundles $(E,F)$ (over $W$) with
their respective gauge connections $A$ and $A'$. In the mixed configurations {\bf GSO}
projection fails to cancel the tachyon. Thus the system is unstable and may flow toward
the annihilation of the brane-antibrane pairs with RR charge for these brane
configurations being conserved in the process.

On the open string sector Chan-Paton factors are $2\times 2$ matrices constructed from
the possible open strings stretched among the different types of branes. Brane-brane and
antibrane-antibrane sectors correspond to the diagonal elements of this matrix.
Off-diagonal elements correspond with the Chan-Paton labels of an oriented open string
starting at a brane and ending at an antibrane and the other one to be the open string
with opposite orientation.

The physical mechanism of brane-antibrane creation or annihilation without violation of
conservation of the total RR charge, leads to consider physically equivalent
configurations of $N_1$ branes and $N_2$ antibranes and the same configuration but with
additional created or annihilated brane-antibrane pairs.

\vskip 1truecm

\noindent
{\it 5.2 D-branes and K-Theory}

The relevant mathematical structure describing the brane-antibrane pairs in general type
I and II superstring theories is as follows:

\begin{enumerate}

\item{} $G_1$ and $G_2$ gauge connections $A$ and $A'$ on the Chan-Paton bundles $E$ and
$F$ over $W$, respectively. Bundles $E$ and $F$ corresponding to branes and antibranes
are topologically equivalent. The groups $G_1$ and $G_2$ are restricted to be unitary
groups for Type II theories and symplectic or orthogonal groups for Type I theories.

\item{} Tachyon field $T$ can be seen as a section of the tensor product of bundles $E
\otimes F^*$ and its conjugate $\bar{T}$ as a section of $E^* \otimes F$ (where $*$
denotes the dual of the corresponding bundle.)

\item{} Brane-antibrane configurations are described by pairs of gauge bundles $(E,F)$.

\item{} The physical mechanism of brane-antibrane creation or annihilation of a set of
$m$ $9$-branes and $9$-antibranes is described by the same U$(m)$ (for Type II theories)
or SO$(m)$ (for type I theories) gauge bundle $H$. This mechanism is described by the
identification of pairs of gauge bundles $(E,F)$ and $(E\oplus H, F \oplus H)$.
Actually
instead of pairs of gauge bundles one should consider classes of pairs of gauge bundles
$[(E,F)]= [E] - [F]$ identified as above. Thus the brane-antibrane pairs really
determine an element of the K-theory group K$(X)$ of gauge bundles over $X$ and the
brane-antibrane creation or annihilation of pairs is underlying the $K$-theory concept
of {\it stable equivalence} of bundles. For 9-branes, the embedded submanifold
$W$ coincides with $X$ and the thus brane charges take values in K-theory group of $X$.

\end{enumerate}

\vskip .5truecm

Consistency conditions for 9-branes ($p=9$) in Type IIB superstring theory such as
tadpole cancellation implies the equality of the ranks of the structure groups of the
bundles $E$ and $F$. Thus $rk(G_1)=rk(G_2)$. The `virtual dimension' $d$ of an element
$(E,F)$ is defined by $d= rk(G_1) - rk(G_2)$. Thus tadpole cancellation leads to a
description of the theory in terms of pairs of bundles with virtual dimension vanishing,
$d=0$. This is precisely the definition of reduced K-theory $\tilde{\rm K}(X)$. Thus
consistency conditions implies to project the description to reduced K-theory.

In Type I string theory $9-\overline{9}$ pairs are described by a class of pairs $(E,F)$
of SO$(N_1)$ and SO$(N_2)$ gauge bundles over $X$. Creation-annihilation is now
described through the SO$(k)$ bundle $H$ over $X$. In Type I theories tadpole
cancellation condition is $N_1-N_2 =32$. In this case equivalence class of pair bundles
$(E,F)$ determines an element in the {\it real} K-theory group KO$(X)$. Tadpole
cancellation $N_1-N_2=32$, newly turns out into reduced real K-theory group $\tilde{\rm
KO}(X)$.

Type IIA theory involves more subtle. It was argued
by Witten in \cite{ktheory} that configurations of brane-antibrane pairs are classified by the
K-theory group of spacetime with an additional circle space ${\bf S}^1 \times X$.
K-theory group for type IIA configurations is K$({\bf S}^1 \times X)$.

\vskip 1truecm

\noindent
{\it 5.3 Ramond-Ramond  Fields and K-Theory}

Ramond-Ramond  charges are classified according to the K-theory groups. In this subsection we
will review that the proper RR fields follows a similar classification. For details
see the recent  papers  by Witten  \cite{ed} and by Moore and Witten \cite{mw}. 

It can be showed that RR fields do not satisfy the Dirac quantization condition.
Thus for example,

\begin{equation}
\int_{W_p} { G_p \over 2 \pi} \not{\in} {\bf Z}.
\end{equation}
The reason of this is the presence of chiral fermions on the brane. The phase of the fermions
contribute with a gravitational term $\lambda = \int_W {1 \over 16 \pi^2} tr\big(R \wedge R \big)$.
This gives a correction to the Dirac quantization. In trying to  extended it for the 
all RR fields $G_p$ in string theory it is necessary introduce new ideas as the 
notion of quantum self-duality and K-theory. Thus RR fields should be generalized 
in the context of K-Theory and we will see that in fact, they find an appropriate 
description within this context. Similar as the RR charge, the 
RR fields find a natural classification in terms of K-theory.

For self-dual RR fields it is a very difficult to find the quantum partition function. For the scalar field
in two-dimensions
it can be obtained  by summing over only one of the periods of the 2-torus. It is not possible to sum
simultaneously over both periods. This description can be generalized to any higher degree $p$-forms $G_p$.
It can be done by defining a function $\Omega(x)$ for $x$ in the lattice $\{ H^1(\Sigma, {\bf Z}) \}$
of periods such that

\begin{equation}
\Omega (x +y ) = \Omega(x) \Omega(y) (-1)^{(x,y)},
\end{equation}
where $(x,y) \equiv \int x \cup y = \int x \wedge y$. 

The partition function can be constructed easily from these data. One first has to identify the 
corresponding period lattice $\lambda$. After that, find the $\Omega$ function as a 
${\bf Z}$-valued function on $\Lambda$ such that it satisfies Eq. (33). Finally one has to construct
the partition function.

\vskip 1truecm

\noindent
{\it Period Lattice for Ramond-Ramond Fields}

Let $X$ be the spacetime manifold. One could suppose that that period lattice are:
$\oplus_{p \ even} H^p(X; {\bf Z})$ for Type IIA theory and $\oplus_{p \ odd} H^p(X; {\bf Z})$ 
for Type IIB theory. However this are not the right choice since the RR charges and fields
actually take values in K-Theory, just as has been described in the last subsection.
Thus, one can see that the period lattice for Type IIA theory is K$(X)$ and for 
Type IIB it is K$^1(X)$. This is more obvious from the anomalous brane couplings. If 
$X= {\bf R} \times Y$ we have

\begin{equation}
{d G \over 2 \pi} = \delta (Y) \sqrt{\hat{A}(Y)} ch(TX). 
\end{equation}
Hence the period lattice is constructed from $\oplus_{p \ even} H^p(X; {\bf R})$ generated by
$\sqrt{\hat{A}} ch(TX)$ for $x=(E,F) \in K(X)$. Still it is necessary to quantize the
lattice by finding the $\Omega$ function and its corresponding quantum partition function. 

\vskip 1truecm

\noindent
{\it The $\Omega$ Function}

In K-Theory there exist a natural definition of the $\Omega$ function given by the 
index theory

\begin{equation}
(x,y) = {\rm Index \ of \ Dirac \ Operator \ on \ X \ with \ values \ in \ x \otimes \bar{y}}\\
= \int_X \hat{A}(X) ch(x) ch(\bar{y}).
\end{equation}
Thus the $\Omega$ function can be defined as 

\begin{equation}
\Omega(x) = (-1)^{j(x)},
\end{equation}
where $j(x)$ is given by the mod 2 index of the Dirac operator with values in the real
bundle $x \otimes \bar{x}$. It can be shown that this definition of $\Omega(x)$
satisfies the relation (33). From this one can construct a quantum partition function
which is compatible with $(i)$ T-duality, $(ii)$ Self-duality of RR fields, $(iii)$
the interpretation of RR fields in K-theory and $(iv)$ description of the brane anomalies.

\vskip 2truecm
\section{String Theory and Noncommutative Gauge Theory}

\noindent
{\it 6.1 Noncommutative D-branes From String Interactions}

Finally in this section we describe briefly some new developments on
the relation between string theory and Connes's noncommutative Yang-Mills theory
(for a survey on noncommutative geometry see the classic Connes
book \cite{alan}). We do not pretend to be
exhaustive but only to remark the key points
of the recent exciting developments \cite{cds,schomerus,sw} (for a nice review
see \cite{douglas}). 

The roughly idea consists from the description of a string propagating in a flat background (spacetime) 
of metric $g_{ij}$ and a NS constant $B$-field $B_{ij}$. The action is given by

\begin{equation} 
{\cal L} = { 1\over 4 \pi \alpha '} \int_{D} d^2 \sigma \  \bigg(g_{ij} \partial_a X^i
\partial^a
X^j  - 2 \pi i \alpha ' B_{ij} \varepsilon^{ab} \partial_a X^i \partial_b X^j \bigg)
\end{equation}
where $D$ is the disc. Or equivalently

$${\cal L} = { 1\over 4 \pi \alpha '} \int_{D} d^2 \sigma \  g_{ij} \partial_a X^i
\partial^a
X^j  - {i \over 2} \int_{\partial D} d \tau  B_{ij} X^i  \partial_{\tau} X^j 
$$
Equations of motion from this action are subjected to the boundary condition
\begin{equation}
 g_{ij}
\partial_n X^j + 2 \pi i \alpha ' B_{ij}
\partial_t X^j |_{\partial \Sigma} = 0.
\end{equation} 

The propagator of open string vertex operators is given by

\begin{equation}
\langle X^i(\tau) X^j(\tau ') \rangle =- \alpha ' G^{ij} \log (\tau - \tau ' )^2 + {i
\over 2}
\Theta^{ij} \varepsilon (\tau - \tau ') 
\end{equation}
where

\begin{equation} 
G^{ij} = \bigg( { 1 \over g + 2 \pi \alpha ' B} \bigg)^{ij}_{S}, \ \ \ \ \ 
\Theta^{ij} = 2 \pi
\alpha ' \bigg( {1 \over g + 2 \pi \alpha ' B}\bigg)^{ij}_{A}.
\end{equation}
Here $S$ and $A$ stands for the symmetric and antisymmetric part of the involved matrix, 
and the logarithmic term determines the anomalous dimensions as usual. Thus $G_{ij}$ is
the effective metric seen by the open strings. While, as was suggested by Schomerus
\cite{schomerus}, the
antisymmetric part $\Theta^{ij}$ determines the {\it noncommutativity}. 

The product of tachyon vertex operators $exp(i p \cdot X)$ and $exp(i q \cdot X)$ for
$\tau > \tau '$ in the short distance singularity is written as

\begin{equation}
exp \bigg( i p \cdot X \bigg) (\tau) 
exp \bigg( i q \cdot X \bigg)(\tau ')  
\sim (\tau - {{\tau}'})^{2 \alpha ' G^{ij}p_iq_j} \cdot exp \bigg( -{1 \over 2}
\Theta^{ij}p_i q_j
\bigg) exp \bigg( i (p+q)\cdot X \bigg)(\tau ') + \dots
\end{equation}
or

\begin{equation}
exp \bigg( i p \cdot X \bigg) * exp \bigg( i q \cdot X \bigg) \sim
exp \bigg( i p \cdot X \bigg) * exp \bigg( i q \cdot X \bigg)
\equiv exp \bigg( {i\over 2} \Theta^{ij} p_i q_j\bigg)exp \bigg( i (p+q) \cdot
X \bigg) 
\end{equation}
where $*$ is defined for any smooth functions $F$ and $G$ over $X$ and it is given by

\begin{equation}
F * G = exp \bigg( {i\hbar \over 2} \Theta^{ij} {\partial \over \partial u_i}
{\partial \over \partial v_j}\bigg) F(x +u) G(x+y). 
\end{equation}
Here the operation $*$ is associative $F*(G*H) = (F*G)*H$ and noncommutative
$F*G \not= G*F$. The above product can be written as $F*G = FG + i\{ F,G\} + \dots  $
where $\{F,G\}$ is the Poisson bracket given by $ \Theta^{ij} \partial_i F \partial_j G.
$ $\Theta$ is determined in terms of $B$. Its give
an associative and noncommutative algebra. In the limit $\alpha ' \to 0$
(ignoring the anomalous dimensions of open string sector) the product of vertex
operators turns out to be the Moyal product of functions on
the spacetime $X$.

Now one can consider scattering amplitude (parametrized by $G$ and $\Theta$) of $k$
gauge bosons of momenta $p_i$, polarizations $\varepsilon_i$ and Chan-Paton wave
functions $\lambda_i,$ $\i = 1, \dots , k$

\begin{equation}
A(\lambda_i, \varepsilon_i,p_i)_{G, \Theta} = Tr \bigg( \lambda_1 
\lambda_2  \dots \lambda_k\bigg) \int d \tau '_i \langle
\prod_{i=1}^k \varepsilon_i \cdot {dX \over d \tau} exp \bigg( i p_i \cdot X \bigg)(\tau '_i)
\rangle_{G, \Theta}. 
\end{equation}

The $\Theta$ dependence come from the factor $exp \bigg( -{i \over 2} \sum_{s >r} p^{(s)}_i
p^{(r)}_j \Theta^{ij}\bigg)$. Thus amplitude factorizes as
$A(\lambda_i, \varepsilon_i,p_i)_{G, \Theta=0} \cdot exp \bigg( - {i \over 2} \sum_{s >r}
p^{(s)}_i \Theta^{ij} p^{(r)}_j \varepsilon(\tau_r - \tau_s) \bigg)$ which depends only
on the cycle ordering of the points $\tau_1, \dots , \tau_k$ on the boundary of the disc
$\partial D.$

For $B=0$ the effective action is obtained under the assumption that the divergences are
regularized through the Pauli-Villars procedure and it is given by

\begin{equation}
S_G = {1 \over g_{st}} \int d^n x \sqrt{G}  \bigg( TrF_{ij} F^{ij} + \alpha ' \ 
{\rm corrections}\bigg). 
\end{equation}
The important case of the effective theory  when $\Theta \not=0$ is incorporated through
the phase factor
and thus one have to replace the ordinary multiplication of wave functions by the $*$
product (effective action is computed by using point splitting regularization)

\begin{equation}
\hat{S}_G = {1 \over g_{st}} \int d^n x \sqrt{G} G^{i i'} G^{j j'}
 \bigg(Tr \hat{F}_{ii'} * \hat{F}^{jj'} + \alpha ' \ 
{\rm corrections}\bigg),
\end{equation}
where $ \hat{F}_{ij} = \partial_i \hat{A}_j - \partial_j \hat{A}_i - i \{\hat{A}_i, 
\hat{A}_j\}_M$ is the noncommutative field strength. Here 
$ \{ F, G \}_M \equiv F*G- G*F$. Thus we get a noncommutative Yang-Mills theory as the
$\Theta$ (or $B$) dependence of the
effective action to all orders in $\alpha '$. Gauge field transformation $ (\hat{\lambda}
* \hat{A})_{ij} = \hat{\lambda}_{ik} * A^k_j$ and 
$ \delta \hat{A}_i = \partial_i \hat{\lambda} + i \hat{\lambda} * \hat{A}_i 
- i \hat{A}_i * \hat{\lambda}.$

For the low varying fields the effective action is given by the Born-Infeld-Dirac
action

\begin{equation}
S = {1 \over g_{st}(\alpha ')^2} \int d^n x \sqrt{ {\rm det} \big( g + \alpha '
(F+B)\big)}.
\end{equation}

The same effective action is described by noncommutative Yang-Mills theory but also by
the standard Yang-Mills theory. They differ only in the regularization prescription. For
the standard commutative case it is the Pauli-Villars one, while for the noncommutative
case it is the point splitting prescription. The two frameworks are equivalent and thus
there is a redefinition of the variable fileds and it can be seen `as a transformation
connecting standard and noncommutative descriptions. The
change of variables known as the {\it Seiberg-Witten map} is as follows

$$ 
\hat{A}_i = A_i - {1\over 4} \Theta^{kl} \{ A_k, \partial_l A_i + F_{li} \} + O(\Theta^2) 
$$

\begin{equation}
\hat{\lambda} = \lambda+ {1\over 4} \Theta^{kl} \{ \partial_l \lambda, + A_j \} +
O(\Theta^2). 
\end{equation}

\newpage

\vskip 1truecm

\noindent
{\it 6.2 String Theory and Deformation Quantization}

Very recently a renewed deal of excitation has been
taken place in deformation quantization theory \cite{bayen}, since the
Kontsevich's seminal paper \cite{k}. In this paper Kontsevich proved by
construction the existence of a star-product for any finite dimensional
Poisson manifold. His construction is based on his more general statement
known as the ``formality conjecture''. The existence of such a
star-product determines the existence of a deformation quantization for
any Poisson manifold. Kontsevich's proof was strongly motivated by some   
perturbative issues of string theory and topological gravity in
two-dimensions, such as, matrix models, the triangulation of the moduli
space of Riemann surfaces and mirror symmetry.

One of the main lessons of the stringly \cite{k} and D-brane \cite{schomerus} 
descriptions of Kontsevich's formula in that of the deformation
quantization for any Poisson manifold requires necessarily of string
theory. In addition this was confirmed in \cite{sw}.
The deformation parameter of this quantization is
precisely the
string scale $\alpha '$ (or the string coupling constant)  which in the
limit $\alpha ' \to 0$ it reproduces the field theory limit but in this
limit the deformation quantization does not exist. The deformation arising
precisely when $\alpha ' \not= 0$ is an indication that deformation
quantization is an stringly phenomenon. Actually it was already suspected
since the origin of the formality conjecture where several mathematical
ingredients of string theory were present.

String action in a background NS constant $B$ field is

\begin{equation}
 S = {1 \over 4 \pi \alpha '} \int_{D} d^2z \partial_a X^i \partial_a X^j
G_{ij} + {1 \over 4 \pi \alpha '} \int_{D} dz d \bar{z} J^i(z) \bar{J}^j(\bar{z}) B_{ij},
\end{equation}
where $J^i(x) = 2i \partial X^i(z,\bar{z})$ and $\bar{J}^i(x) = 2i \bar{\partial} X^i(z,\bar{z})$.

Define the function

\begin{equation}
F(X(x)) = V[F](x) := {1 \over (2 \pi)^{d/2}} \int d^dk \hat{F}(k) V_k(x)
\end{equation}
where $V_k(x)= : exp\big(i k_i X^i(x)\big) :$ is the vertex operator. OPE between
$J's$ and $V's$ operators leads to $V[F](1) V[G](0) \sim V[FG](0) + \dots  $. The
introduction of a NS constant $B$ field in the action `deforms' the OPE leading to

\begin{equation}
(V[F](1) V[G](0)^B \sim V[F*G](0) + \dots  
\end{equation}
where the $*$ product will be determined. It can be obtained by computing the
$N$-point correlations functions for the complete action (49) (including the $B$-term)

$$
\langle \Phi_1 \Phi_2 \dots  \Phi_N \rangle^B = {1 \over Z} \langle  \Phi_1 \Phi_2 \dots  \Phi_N 
exp \bigg( -{1 \over 4 \pi \alpha '}\int_{\cal H}  dz d \bar{z} J^i(z) \bar{J}^j(\bar{z}) B_{ij}
\bigg) \rangle^B
$$

\begin{equation}
= {1 \over Z} \sum_{n=0}^{\infty} \big(- {1 \over 4 \pi \alpha '} \big) {1 \over n!}
\int_{{\cal H}_n^{\varepsilon}} d^d z d^d \bar{z} \langle \Phi_1 \Phi_2 \dots \Phi_N
\prod_{a=1}^n B_{i_a j_a} J^{i_a}(z_a) \bar{J}^{i_a}(\bar{z}_a) \rangle
\end{equation}
where $Z:= \langle exp \bigg(  -{1 \over 4 \pi \alpha '}\int_{\cal H}  dz d \bar{z} J^i(z)
\bar{J}^j(\bar{z}) B_{ij} 
\bigg) \rangle^B$ and ${\cal H}_n^{\varepsilon}:= \{ (z_1,z_2, \dots, z_n) | Im (z_a) > \varepsilon,
\ |z_a - z_b| > \varepsilon \ {\rm for} \ a \not= 0 \}$.
We choice $\Phi_1 = V[F](1)$ and $\Phi_2[G](0)$. Now using the usual OPE of the $J's$ and $V_k's$ 
operators and substituting all this at Eq. (52) we get deduce the explicit form fo the $*$ product
and it is given by

\begin{equation}
F*G = \sum_n (4 \pi \alpha ')^n W_n B_n(F,G),
\end{equation}
where $W_n$ are the weight function

\begin{equation}
W_n:= {1 \over (2 \pi)^{2n}}{1 \over n!} \int d^nz d^n \bar{z} \prod_{a=1}^n
\bigg({1 \over z_a -1} {1 \over \bar{z}} - {1 \over \bar{z} -1}{1 \over z_a} \bigg)
\end{equation}
and $B_n(F,G)$ are bi-differential operators

\begin{equation}
B_n(F,G):= \sum \Theta^{i_1 j_1} \Theta^{i_2 j_2} \dots  \Theta^{i_n j_n}
\partial_{i_1} \partial_{i_2} \dots \partial_{i_n} F \ \partial_{j_1} \partial_{j_2}
\dots \partial_{j_n} G,
\end{equation}
where $\Theta^{ij}$ is that given in Eq. (40).

This is precisely the formula given by Kontsevich in \cite{k} for the $*$ product on any Poisson
manifold. In this case the Poisson manifold is the spacetime $X$ and $\Theta_{ij}$ is the Poisson-like
structure.

\newpage

\vskip 2truecm

\centerline{\bf Acknowledgements}
We are very grateful to the organizing committee of the {\it Third
Workshop on Gravitation and Mathematical Physics} for the hospitality. One
of us `O. L-B. is supported by a CONACyT graduate fellowship.

\vskip 2truecm


\end{document}